\documentclass[sigconf,natbib=true]{acmart}
\AtBeginDocument{%
  \providecommand\BibTeX{{%
    \normalfont B\kern-0.5em{\scshape i\kern-0.25em b}\kern-0.8em\TeX}}}

\setcopyright{none}
\acmYear{2024}
\acmDOI{}

\acmConference[EAI-KDD'2024]{Make sure to enter the correct
  conference title from your rights confirmation emai}{August 25--25,
  2024}{Barcelona, Spain}
\acmPrice{}
\acmISBN{}

\usepackage[linesnumbered,ruled]{algorithm2e}
\usepackage{multirow}
\usepackage{bm}
\usepackage{bbm}
\usepackage{amsfonts}
\usepackage{tikz}
\newcommand{\ind}{\mathbbm{1}}

\usepackage{amsmath}

\def\1{\bm{1}}

\def\mD{{\bm{D}}}

\def\mW{{\bm{W}}}

\DeclareMathAlphabet{\mathsfit}{\encodingdefault}{\sfdefault}{m}{sl}
\SetMathAlphabet{\mathsfit}{bold}{\encodingdefault}{\sfdefault}{bx}{n}

\def\gI{{\mathcal{I}}}

\def\gL{{\mathcal{L}}}

\def\gU{{\mathcal{U}}}

\newcommand{\R}{\mathbb{R}}

\usepackage{tikz}
\usepackage{natbib}
\usepackage{caption}
\usepackage{subcaption}
\usepackage{wrapfig}
\usepackage{url}

\begin{document}

%\title{Tails or Outliers? an Adversarial Approach on Semantic Tails of Disadvantaged Items}
\title{POSIT: Promotion of Semantic Item Tail via Adversarial Learning}
%An Adversarial Technique for Promoting Semantically Disadvantaged Items}


\author{Qiuling Xu}
\email{qiulingx@netflix.com}
\orcid{0000-0001-7399-4577}
\affiliation{%
  \institution{Netflix Research}
  \city{Los Gatos}
  \state{CA}
  \country{USA}
  \postcode{95008}
}

\author{Pannaga Shivaswamy}
\email{pannaga.datta@gmail.com}
\affiliation{%
  \institution{Pocket FM}
  \city{Bengaluru}
  \country{India}}

\author{Xiangyu Zhang}
\email{xyzhang@purdue.cs.edu}
\affiliation{%
  \institution{Purdue University}
  \city{West Lafayette, IN}
  \country{USA}
}

\newcommand{\system}{{POSIT }}

\begin{abstract}
In many recommendations, a handful of popular items (e.g., movies / television shows, news, etc.) can be dominant in recommendations for many users. However, we know that in a large catalog of items, users are likely interested in more than what is popular. The dominance of popular items may mean that users will not see items that they would probably enjoy. In this paper, we propose a technique to overcome this problem using adversarial machine learning. We define a metric to translate the user-level utility metric in terms of an advantage/disadvantage over items. We subsequently used that metric in an adversarial learning framework to systematically promote disadvantaged items. Distinctly, our method integrates a small-capacity model to produce semantically meaningful weights, leading to an algorithm that identifies and promotes a semantically similar item within the learning process. In the empirical study, we evaluated the proposed technique on three publicly available datasets and seven competitive baselines. The result shows that our proposed method not only improves the coverage, but also, surprisingly, improves the overall performance. \footnote{Code is available at \url{https://github.com/qiulingxu/POSIT}}
\end{abstract}

\begin{CCSXML}
<ccs2012>
   <concept>
       <concept_id>10002951.10003317.10003338.10003345</concept_id>
       <concept_desc>Information systems~Information retrieval diversity</concept_desc>
       <concept_significance>500</concept_significance>
       </concept>
 </ccs2012>
\end{CCSXML}

\ccsdesc[500]{Information systems~Information retrieval diversity}

\keywords{recommendation, coverage, diversity, adversarial learning}


\iffalse
\begin{teaserfigure}
  \includegraphics[width=\textwidth]{sampleteaser}
  \caption{Seattle Mariners at Spring Training, 2010.}
  \Description{Enjoying the baseball game from the third-base
  seats. Ichiro Suzuki preparing to bat.}
  \label{fig:teaser}
\end{teaserfigure}
\fi 

\newcommand{\todoc}[2]{{\textcolor{#1}{\textbf{#2}}}}
\newcommand{\todored}[1]{{\todoc{red}{\textbf{[#1]}}}}
\newcommand{\todogreen}[1]{\todoc{green}{\textbf{[#1]}}}
\newcommand{\todoblue}[1]{\todoc{blue}{\textbf{[#1]}}}
\newcommand{\todoorange}[1]{\todoc{orange}{\textbf{[#1]}}}
\newcommand{\todobrown}[1]{\todoc{brown}{\textbf{[#1]}}}
\newcommand{\todogray}[1]{\todoc{gray}{\textbf{[#1]}}}
\newcommand{\todopink}[1]{\todoc{pink}{\textbf{[#1]}}}
\newcommand{\todopurple}[1]{\todoc{purple}{\textbf{[#1]}}}
\newcommand{\revision}[1]{{#1}}
\newcommand\TODO[1]{\textcolor{red}{#1}}
\newcommand{\ql}[1]{\todoblue{QL: #1}}
\newcommand{\ps}[1]{\todoblue{Pannaga: #1}}
\newcommand*\circled[1]{\raisebox{.5pt}{\textcircled{\raisebox{-.9pt} {#1}}}}

\maketitle

\section{Introduction}

Recommender systems are used extensively in many consumer web applications such as streaming services \cite{Gomez-Uribe2015}, video recommendations \cite{youtube}, news feed recommendations \cite{10.1145/2835776.2835848, Agarwal2014}, etc. The primary goal of  recommender systems is to recommend appropriate {\em items} 
%such that they provide some utility to the users 
such that users are engaged, entertained or feel connected.  The catalog of items can be movies, TV shows, news articles, videos, merchandise etc. Depending on how the training data is collected and how a model training is done many different types of biases can exist in a recommender system \cite{10.1007/s10791-017-9312-z,chen2021bias, 10.1145/3109859.3109912}; overcoming such biases is an important research direction in the field of recommender systems and many approaches have been proposed to go beyond accuracy and address some of these biases (e.g., \cite{10.1145/2926720, 10.1145/1864708.1864761}).

 In this paper, we focus mainly on the popularity bias \cite{popularity-bias, popularity-bias-reranking}, which is a particular type of bias where a recommender system recommends many popular items at a possible disadvantage to many other relevant items. The unpopular items, while have less user interaction, constitute the majority of the catalog and are typically referred to as the long tail of the catalog.  Promoting such infrequent items from the long tail is potentially crucial to users' satisfaction and a higher chance of success to every item. Unlike popular items, these infrequent items can coincide with personalized taste, can provide an unexpected experience other than mainstream popularity, motivate users to explore deeper into the catalog, and evoke a sense of freshness~\cite{unexpected}. On the supply side of items, many more items have a chance to succeed on the platform such that a handful of minority items do not suppress the chance of a vast majority of items.

\iffalse
\begin{figure}[t]
\includegraphics[width=0.45\textwidth]{figures/Motivation.pdf}
\caption{In figure (A), we compare the coverage ratio  of unique recommended items divided by the upper-bound (see Equation~\ref{eq:converage_ratio_def}). It reflects the coverage of unique and rare items, as well as inter-user diversity. In figure (B), we measure the performance of the recommendation on the item side (see Equation~\ref{eq:itemrecall}). In figure (C), we measure the performance on the user side. The experiments are carried out on the MovieLens 20 million dataset}
\label{fig:motivation}
\vspace{-8pt}
\end{figure}

 Figure~\ref{fig:motivation}(A) illustrates the concentration phenomenon of popular items on the MovieLens~\citep{movielens} dataset. We measure how many unique items are recommended at the top compared to the upper limit of how many items can be recommended. A rigorous definition can be found in eq.(\ref{eq:converage_ratio_def}). It shows that a strong baseline only covers about 15\% items while our proposed technique can significantly improve coverage while not reducing overall performance.
\fi

To alleviate this problem of a few items being concentrated at the top of the recommendation, an intuitive solution is to adjust the weight of items during learning. The weights are adjusted so that the most frequent items are lowered while the rare ones are increased during training. Although this practice may alleviate the problem to some extent, there are many challenges associated with it. First, we need to clearly define an appropriate weight for each item, while controlling how much we weigh one over the other. Second, it is unclear whether the frequency of the item alone should play such a big role in up-weighing an item, or whether other factors should be considered.

Any technique that reduces popularity bias and increases item coverage has the risk of reducing the overall performance of the recommender system. By promoting many more less-frequent items, it is possible that the users would receive even less utility compared to the more popular counterparts that could have been recommended. %Figure~\ref{fig:motivation}(C) illustrates this trade-off. Observe that the performance of some baselines are quite affected with increasing coverage.  
In this paper, we propose a method to decrease popularity bias, increase coverage, and, surprisingly, {\em also improve the overall performance of recommendation systems}. Rather than directly optimizing the tail of the catalog, our technique focuses on the ``semantically meaningful'' tail. A semantically meaningful tail is a group of similar items in the tail that exclude potential outliers. Unpopular items on the tail can often be outliers. Excluding these outliers and only keeping semantically similar items helps to improve generalization.

\begin{figure*}
\includegraphics[width=.95\linewidth]{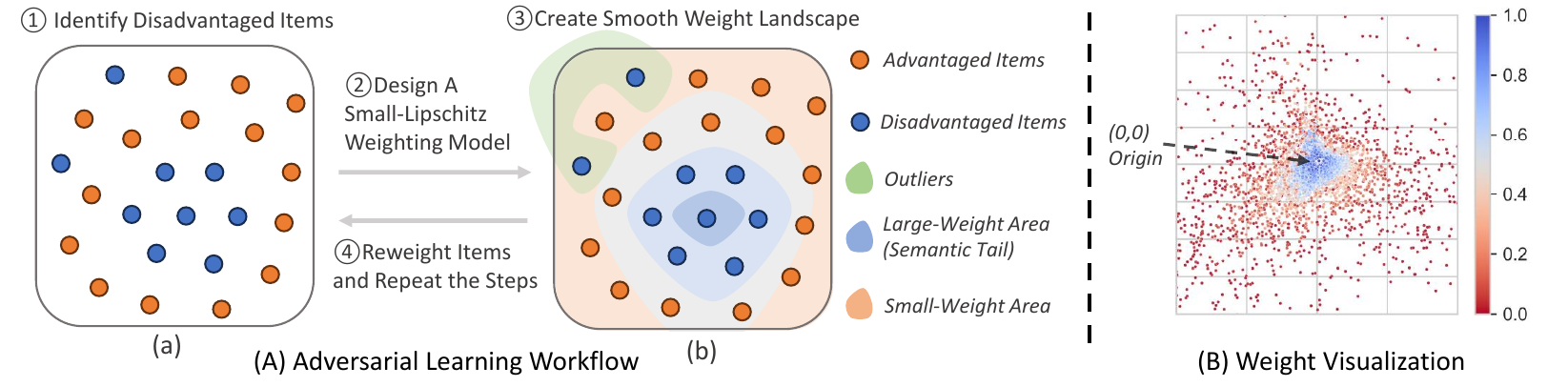}
\vskip -0.1in
\caption{Figure (A) presents the adversarial learning workflow. In step \circled{1}, item advantage is quantified using an itemwise metric adapted from Recall (Eq. \ref{eq:advantage_score_with_popularity}). Items are visually distinguished as blue (disadvantaged) and orange (advantaged) for illustrative clarity.  In steps \circled{2} and \circled{3}, an adversarial model,  constrained by a small Lipschitz constant, assigns a continuum of weights to items. Driven by adversarial optimization, large weights are assigned to disadvantaged items, while small weights are assigned to advantaged ones. As this model produces smooth weight landscapes, to maximize the loss, these assigned weights naturally focus on the disadvantaged clusters while filtering out outliers. This process leads to the formation of \textit{semantic tails}, a term that refers to clusters of disadvantaged but semantically related items. Step \circled{4} involves adjusting the weights of these items and iterating the process. As weights are reassigned, the semantic tails dynamically change. We repeat these steps and continuously track the semantic tails via the adversarial model. Figure (B) presents a visualization of semantic tails on MovieLens, based on Principal Component Analysis (PCA). Each dot represents an item, with the color indicating its associated weight.  Rare items appear near the origin in this representation. Observe semantic tails proximal to the center, characterized by a gradual decrease in weights extending outward. More comparison and details can be found in \autoref{sec:visualization}.}
\label{fig:visualize_weight}
\vskip -0.1in
\end{figure*}

An example is illustrated in \autoref{fig:visualize_weight}(A.b). In this figure, each point visualizes an item, with blue being the unpopular ones (disadvantaged) and orange indicating the others. The colored area indicates assigned weights. Items in the blue area are assigned large weights and promoted. Observe that the unpopular items in the deep blue area are close to each other. This cluster of semantically similar items in the tail forms a semantic tail. Our method tracks these tails and assigns large weights to them. In contrast, there exist two distant unpopular items in the green areas -- these are outliers. Constrained by the capacity of the adversarial model, these outliers receive small weights similar to those of their neighbors.

\autoref{fig:visualize_weight}(B) effectively echoes our theoretical concepts. It represents MovieLens movies after PCA reduction, placing less popular items near the origin due to sparse interactions. Each item is color-coded based on weights from the adversarial model, with the blue region highlighting semantic tails and the promotion of less popular movies in our approach. This smooth distribution of weights confirms our analysis, demonstrating that similar items are assigned comparable weights. For additional information, see \autoref{sec:visualization}.

%Visualization of MovieLens in Figure~\ref{fig:visualize_weight}.(B) further supports our claim.  Each point visualizes a movie after dimension reduction (Principle Component Analysis) over the user-item interaction matrix. As unpopular items receive sparse interactions, they are rendered close to the origin (0,0) during the PCA. The color indicates the weights assigned by the adversarial model. Observe that the captured semantic tails cover unpopular movies near the origin, which are promoted during learning. The smooth weight landscape observed is consistent with our analysis. Note that many unpopular outliers exist and are assigned small weights after applying adversarial models. Due to the limited space, please refer to Appendix~\ref{visualization} for more details.
%Deep red region are denser and therefore indicates the majority of movies. The deep blue region are sparse and therefore indicates the tail. 
% The semantically meaningful tail are the movies between these two regions. These implicitly identified disadvantaged items are subsequently promoted through learning. %Figure~\ref{fig:motivation}(B) shows the large increase of performance on the tail by focusing on ``semantically meaningful'' tail. 

The main contributions of our work are as follows. First, we identify that typically in a recommender system we care about user-level ranking metrics, but for increasing item coverage, we need to bridge the gap and make sense at an item level. We thus start with a user-level metric and show how we can identify which items are at an advantage/disadvantage. Second, we use the defined metric in an adversarial model learning setting such that items at a disadvantage are promoted. Due to the smoothness of the adversarial model, much more semantically meaningful items are promoted rather than extreme outliers. Finally, we show empirical results where we compare the proposed approach with several other baselines. The experiments show that while 
%we significantly improve the coverage of items, we do not lose any performance. Indeed, 
we significantly increase the coverage of items, we also improve the overall performance of recommendations. Finally, we show a number of empirical observations to shed light on how the proposed technique is able to achieve both accuracy and coverage.

The remainder of the paper is organized as follows. We review other related work in Section \ref{sec:related}. We define the problem setting and introduce the notation in \ref{sec:setting}. We define a metric to define the advantage of an item and further use it in an adversarial learning setting to optimize it in Section \ref{sec:adversarial}. We briefly introduce those baselines compared in Section~\ref{sec:baselines}. We show our empirical results on three publicly available datasets in Section \ref{sec:expts} and conclude in Section \ref{sec:conclude}.

\iffalse
\ql{Structure of the paper
1. Introduce Item Diversity Concentration phenomenon. 

2. The related works, Weight rebalancing. Reranking. Tail Optimization.

3. Unique challenges in Title Analysis

4. Design of Adversary

5. proper choice of metric}

Many of such services has the In this retail model, the recommendation system help business distribute items like  products or services to the desirable users. In this paper, we focuses on a subgroup of businesses which distributes movies and musics.  Different from general e-commerce, the number of provided movies and musics are orders of magnitude less than the users.

It is known that the recommendation system has the concentration phenomenon.  The system will favor popular and trendy items compared to an average item. This popularity bias is prone to enlarge during learning the recommendation system \cite{mcauley_2022}. 

\fi
\section{Related Work}
\label{sec:related}

Adversarial machine learning has recently gained popularity \cite{biggio2018wild, 2022_CVPR, DBLP:journals/corr/abs-1812-04948} in the literature. Some of these works, such as \cite{DBLP:journals/corr/abs-1808-03908}, have explored the security aspects of recommendation systems. Unlike these studies, our research focuses on improving key metrics without factoring in potential attacks. More recently, an instance-level weight-based adversary was proposed in \cite{arl} the context of supervised learning and was further studied in the context of recommender systems \cite{adv_friend}. The work in \cite{adv_friend} improves model performance in users where the baseline models had difficulty. In addition, there have been other robust optimization approaches \cite{10.1145/3485447.3512255} to improve the recommender system from a user performance perspective. However, to the best of our knowledge, our work constitutes a first attempt to study an adversarial machine learning problem from the item side on the recommender system.

Popularity bias is a popular topic in the recommender systems literature. There have been existing works on handling popularity bias \cite{IPW,adomavicius2009toward,rockafellar2000optimization}. Many of these approaches require identifying a priori which items are in the ``long tail" and which ones are in the short tail of recommendation. It is difficult for their manual design process to address the popularity bias problem in a systematic way. Compared to those approaches, we do not need to a priori identify which items are in the long-tail versus short-tail category.  We compare two competitive baselines in this category, including IPW~\cite{IPW} and CVaR~\cite{rockafellar2000optimization} in Figure~\ref{fig:visualize_weight} and experiments. The results show our significant improvement in addressing the popularity bias problem. 

Many approaches have been proposed for handling popularity bias or ensuring a degree of fairness among items  in various studies~\cite{adomavicius2009toward,10.1145/3109859.3109912,popularity-bias-reranking, singh2018fairness, 10.1145/3471158.3472260,post, 10.1145/3306618.3314309}. These methods typically focus on post-processing rather than altering the core learning algorithm. During post-processing, they rerank the recommendations based on additional factors like frequency, exposure etc. However, these methods involve a direct trade-off between performance and coverage, while our method is focused on improving the underling model itself. Our method is able to alleviate popularity bias without sacrificing the performance. Moreover, the post-processing methods are complementary to ours in that they can be applied on top of any learned model.  In our experiments, we chose two representative baselines Rerank~\cite{popularity-bias-reranking} and post-processing~\cite{post} for comparison. 

Some methods~\cite{causal2,causal1} use causality to reduce popularity bias. They estimate the causal effect of popularity and mitigate this effect. However, these methods require that the model's output and loss meet some probabilistic interpretations. We found that this requirement can be too strong and will reduce the model performance for some competitive models. Differently, our method calibrates the weight of loss through adversary, without the assumption of loss format, it can work across a wider class of models. We include one of the latest baselines~\cite{causal2} for evaluation.

There is also work directly addressing producer-side fairness ~\cite{10.1145/3471158.3472260} in two-sided marketplaces. However, their framework requires additional information about the producers for item modeling. Differently,  we leverage the user-side utility and convert that into an item-side score and use it in an adversarial learning framework. This practice eliminates the need for additional producer information.

\section{Problem Setting and Preliminaries}
\label{sec:setting}

In this paper, we focus on recommendations in an implicit feedback setting. In this setting, we collect users' interactions with a catalog of items. As a common assumption, the interaction of the users indicates their preference. Unlike the explicit setting, the users will not directly express their preferences like a rating or score. The less strict assumption of implicit setting makes it important and popular as it closely aligns with many real-world applications, where a majority of users do not give explicit feedback.

Consider a training dataset consisting of $|\gI|$ items and $|\gU|$ users, a user-item interaction matrix $\mD \in \mathbf{R}^{|\gU| \times |\gI|}$. User $u$ has the interaction history $\mD_u \in \{0,1\}^{|\gI|}$, denoting that the user $u$ has interacted with the items that are ones. When $D_{ui}=0$, it indicates that the user $u$ did not interact with the item $i$.  Such interactions can include a click, a buy, or watching a movie. The goal is to learn a recommendation model $f: \{0,1\}^{|\gI|} \rightarrow \mathbf{R}^{|\gI|}$. The model gives a score for every item that can be used for the item ranking.  Given a trained model, the goal then is to take a test user's historically interacted items and rank the non-interacted items that are relevant to the user.

To evaluate the performance of the recommendation system $f$, we adopt a widely used metric - "Recall@k". We also consider widely used metrics such as normalized discounted cumulative gain (NDCG) \cite{Manning_etal_2008}. Since these two metrics were highly correlated with each other in our experiments, we mainly focussed on the recall. Consider the preference for the item $i$ for the user $u$ as $D_{ui}$. The recommendation system scores each item via $f$ and ranks them by their scores for a particular user in descending order. Let the rank of the item $i$ for the user $u$ be $r_{ui}$. The Recall@K is defined as 
\begin{equation}    
\mathrm{Recall@}k = \frac{1}{|\gU|} \sum_{u \in \gU} \frac{\sum_{i \in \gI} \ind\left[D_{ui}=1 \land r_{ui} \leq k \right]}{\min(k,\sum_{i \in \gI} D_{ui})} \, .
\end{equation}
Intuitively, it calculates the fraction of items within the top $k$ ranked items that are actually relevant for  user $i$.

In addition to recall, we use coverage to measure how well the recommendation system can cover the entire catalog of items. Consider a batch of 100 randomly selected users $\gU$.  Coverage is defined as the unique number of items included in the top $k$ recommendations for these 100 users. \revision{Note that we fix the batch size 100 for fair comparisons. Choosing a larger batch size results in a similar observation, as shown in Appendix~\ref{sec:more_ablation}.}
\begin{equation}
    \textrm{Coverage@}k = \left|\bigcup_{u \in \gU} \left\{i \in \gI | r_{ui} < k \right\}\right|
    \label{eq:coverage_def}
\end{equation}
We report this number averaged among batches of 100 users. A higher coverage number indicates that the system can \textit{ cover diversified items for different users}. We also report the ratio of coverage against the theoretical upper-bound. The upper bound is attained when each user is recommended with completely disjoint items. We define the coverage ratio as follows:
\begin{equation}
    \textrm{Coverage Ratio@}k = \frac{\mathrm{Coverage@}k}{\min(k|\gU|, |\gI|)} 
    \label{eq:converage_ratio_def}
\end{equation}

\textbf{Loss.} Before delving into the details, consider a recommendation system $f$ parameterized by $\mW$, and a training loss for user $u$ and item $i$ as  $\gL(D_u;\mW)_i$. The training objective can be described as  
\begin{equation}
    \begin{split}
        \min_{\mW} \sum_{ui} \gL(D_u;\mW)_i
    \end{split}
    \label{eq:general_loss}
\end{equation}

\textbf{EASE.} We consider a particular choice of loss function that was used in EASE~\cite{EASE} which is a strong baseline for recommender tasks. Unlike common matrix factorization approaches, EASE uses an item similarity matrix $\mW \in \mathbf{R}^{|\gI| \times |\gI|}$. However, to avoid learning the trivial solution of the identity matrix, EASE enforces the diagonal of the learned matrix to be $0$.
The optimization can be written as
\begin{align}
    &\min_{\mW} \|D\mW- D\|^2_2  + \lambda \|\mW\|^2_2  
    ~~~~~\ \ \  s.t., \text{diag} (\mW) = 0 \label{eq:EASE}
\end{align}
We adopt EASE as the base recommender because its results on the implicit-feedback setting are very competitive. We will evaluate the performance of different methods applied in EASE. While mainly evaluating based on EASE, our method is general and can be applied to other base recommenders as well. Although at the time of training, a model like EASE (or Matrix Factorization) uses a squared loss, the loss is only a proxy at the time of optimization. The ultimate goal of the model trained is to still perform well on a ranking metric such as NDCG or Recall that we discussed above.

\section{Adversarial Learning Approach}
\label{sec:adversarial}

In this section, we first propose a way to identify items that are at a disadvantage and then propose a technique to promote such items. %A workflow of our approach can be found in Figure~(\ref{fig:workflow}).

\subsection{Identifying Items at a Disadvantage}
As we discussed in the Introduction, a typical recommender system may bias towards certain types of items, like popular items.  In this paper, we take an agnostic approach to such issues and adversarially remedy the problem by identifying semantically similar items potentially at a disadvantage and promoting them.  We will formally measure the disadvantage of an item by a score whose magnitude indicates the magnitude of the disadvantage. The proper choice of the advantage score is critical for the final results. 
%These biases could be caused by various reasons. As one example, the discrepancy between training loss (e.g. $\ell_2$ loss) and evaluation metric (e.g. NDCG or Recall) on a recommender system will make the recommender bias towards the item with a small training loss but with a large evaluation score. It can also be based on how the training data is collected and what policy was used for serving the training data etc. 

 A major challenge to designing an advantage score is the mismatch between the training metric and the evaluation metric. Usually, the training loss function, like the $\ell_2$ loss (e.g. in eq.~\ref{eq:EASE}), is not directly used in evaluating a recommender system, although it may serve as a good proxy at training time.  During the evaluation, the performance of the recommender system is computed by ranking items at the user level. Thus, directly using training loss as the advantage score is not a good choice. Moreover, we empirically find that the disadvantaged items tend to have a small training loss compared to an average item due to the sparsity of the training data. This is, however, the opposite of what an advantage score should indicate. Due to the sparse nature of the dataset, items with fewer interactions will tend to get smaller training losses. We, therefore, focus on taking a user-level ranking metric such as recall and attribute it to item level such that we can easily identify which items are at a disadvantage. In this way, the item-level metric will be directly related to the ultimate metric we care about in the system.

We use a metric derived from Recall@k, which we refer to as ``Item Recall@k''. This concept aims to attribute Recall to each of the items that are part of the recall calculation. As a result, it biases more towards the item tails' performance.  To calculate it, we generate a score matrix $R$ of the same size as $D$ following the rules of Recall where $r_{ui}$ is the rank of item $i$ for user $u$: 
\begin{equation}    
R_{ui} = \ind\left[D_{ui}=1 \land r_{ui} \leq k \right]\, .
\end{equation}
Next, we take the average of the score for each item. To decide the denominator for the average, we can either count the number of all users or the number of users with interactions with that item. The difference between these two options is whether the advantage score should consider popularity bias or not. When using the size of whole users as the denominator, we essentially reflect the popularity bias in the advantage score. Since the denominator is the same for all items, items that would end up in the top $k$ for more users would end up having a higher score. The other option removes the popularity bias, normalizing by the number of interactions for that item. Therefore, Item Recall@$k$ for item $i$ can be written as $S_i$ in two versions,
\begin{align}
    & S_i = \frac{1}{|\gU|} \sum_{u \in \gU} R_{ui} \quad & \textrm{(Reflects popularity bias)} \label{eq:advantage_score_with_popularity}\\
& S_i = \frac{1}{\sum_{u \in \gU} D_{ui}} \sum_{u \in \gU} R_{ui} \quad & \textrm{(Without popularity bias)} \label{eq:advantage_score_wo_popularity}
\end{align}    
When $S_i$ is high in equation (\ref{eq:advantage_score_with_popularity}), that means item $i$ is relevant and would end up in the top $k$ more often showing that the item being at an advantage given a ranking.

In Appendix~\ref{sec:ablation}, we compare these two options and another frequency metric. The results show that using eq.(\ref{eq:advantage_score_with_popularity}) for optimization yields better empirical performance. During the evaluation, different items may have very different sizes of interactions. To reflect the performance on rare items, we include the formula eq.(\ref{eq:advantage_score_wo_popularity}) with uniform item weights. The Item Recall@$k$ is aggregated over all items as the follows,
\begin{equation}    
\mathrm{Item\,Recall@}k = \frac{1}{|\gI|} \sum_{i \in \gI} \frac{\sum_{u \in \gU} \ind\left[D_{ui}=1 \land r_{ui} \leq k \right]}{\sum_{u \in \gU} D_{ui}} \, .
\label{eq:itemrecall}
\end{equation}

\begin{algorithm}[t]
\small
\LinesNumbered
\SetKw{Initial}{Initialize}
\DontPrintSemicolon
 \SetKwInOut{Input}{Input}\SetKwInOut{Output}{Output}
 \Input{Interaction Matrix $D$ subscripted by item $i$ and user $u$, training epochs $T$, number of batches in each epoch $L$, $\ell_2$-regularizer $\lambda$,  momentum $m$.}
\Output{A recommender $f$ and an adversary model $a$}
\BlankLine
\Initial $\hat{S}_i \leftarrow 0$, $\mW \leftarrow 0$, parts of $\phi \leftarrow 0$\;
\For {$t \leftarrow 1$ \KwTo $T$} {
\For {$b \leftarrow 1$ \KwTo $L$} {
    $\hat{U} \subset \gU$ \tcp{Sample a batch of users}
    $\hat{S}_i \leftarrow (1-m)\hat{S}_i + m  S_i$%\frac{\sum_{i \in \hat{U}} \ind\left[D_{ij}=1 \land r_{ij} \leq k \right]}{|\hat{U}|}$ 
    \tcp{Estimate advantage score of item i from a batch}
    $\bar{\alpha}_i =   \frac{|\gI|a(D_{:i};\psi)}{\sum_{i'} a(D_{:{i'}};\psi)}$ \tcp{Normalized weights}
    \textrm{Optimize}\, $\min_{\mW}  \sum_{i \in \gI} \bar{\alpha}_i \sum_{u\in \hat{U}} (D_{u:}\mW- D_{u:})_i^2$ $+ \lambda \|\mW\|^2_2$ \; %\gL_f(f,D_i;\mW)_j$\;  
    $\textrm{diag}(\mW) \leftarrow 0$ \tcp{Constraints of EASE}
    \textrm{Optimize}\, $\max_{\psi} \sum_{i \in \gI} \bar{\alpha}_i (-\hat{S}_i)$
    }
    %\EndFor
\textrm{Evaluate on validation set and save the best model.} \;
\textrm{Update learning rate by a scheduler.}
}
%\EndFor
\caption{The Procedure of \system}
\label{alg:proc}
\end{algorithm}

\subsection{Adversarial Models}

Once we have a score for each item $i$ defined above, we consider the following adversarial reweighted learning formulation. The recommender optimizes a re-weighted loss from the adversary model. The adversary's goal is to identify and promote items at a disadvantage. The adversarial model tracks the semantic tails, where semantic groups of disadvantaged items are implicitly identified, and creates smooth semantic weights based on the advantage scores.  Semantic tails filter outliers and improve generalization during optimization. Formally speaking, let us denotes the adversary model $a: \{0,1\}^{|\gU|} \rightarrow \R$ parameterized by $\psi$, the feature of item $i$ as $D_{:i}$ and the item advantage score of item $i$ defined in eq.(\ref{eq:advantage_score_with_popularity}) as $S_i$, then we have the following formulation:
\begin{align}
    \textrm{Learner}: & \min_{\mW} \sum_{ui} a(D_{:i};\psi) \gL(D_u;\mW)_i \nonumber \\
    \textrm{Adversary}: & \max_{\psi} \sum_{i \in \gI} a(D_{:i};\psi) (-S_i) \, . \label{eq:general}
\end{align}
Additionally, the adversarial model aims to create semantic tails, where items with similar semantics receive similar weights. This is defined by the subsequent mathematical constraint,
\begin{equation}
    D_{:i} \approx D_{:j} \Rightarrow a(D_{:i}) \approx a(D_{:j}) \, .
    \label{eq:semantic_property}
\end{equation}

\begin{figure}[t]
\includegraphics[width=.65\linewidth]{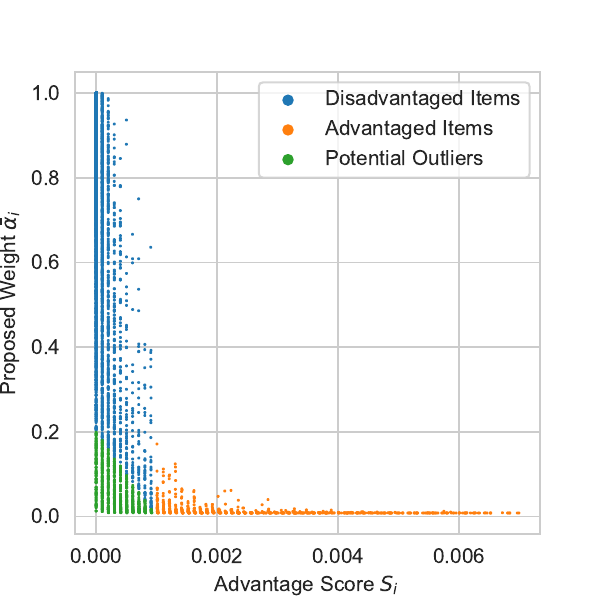}
\caption{Illustration of Outlier Filtering during Re-weighting on Movie-Lens Dataset.  This diagram reveals the relationship between the weights proposed from the adversarial model and the corresponding advantage scores, with each dot representing a specific movie. The X-axis details the advantage score, while the Y-axis outlines the weight.  For a better explanation, we group movies into three groups. It is worth noting that these boundaries are drawn manually and are only for illustration purposes. The blue and orange groups symbolize disadvantaged and advantaged movies, respectively, which are aptly promoted or demoted according to their advantage score. Conversely, the green group highlights potential outliers that, despite being disadvantaged, are not adequately promoted due to the semantic constraints of the adversary, thereby preventing a shift in focus to dissimilar items.}
%The orange one indicates potential outliers. Though being disadvantaged, they are not properly promoted. This is because of the semantic constraints of adversary prevents shifting focus to dissimilar items.}
\label{fig:outliers}
\vspace{-14pt}
\end{figure}

%We show that this property can be fulfilled through a small-Lipschitz neural network. The Lipschitz constant indicates the smoothness of model. A smooth model tends to have similar outputs when inputs are similar. Mathematically a neural network of Lipschitz 
We demonstrate that the desired property can be achieved by employing a small-Lipschitz neural network. The Lipschitz constant is a measure that reflects the smoothness of the model, where a smooth model is characterized by having similar outputs for similar inputs. Specifically, a neural network with a Lipschitz constant of $k$ under the $\ell_2$-norm satisfies the following relationship.
\begin{equation}
 \forall \|D_{:i} - D_{:j}\|_2 \leq \epsilon \Rightarrow| a(D_{:i}) - a(D_{:j})| \leq  k\epsilon
\end{equation}
This suggests that the adversary can maintain semantically meaningful weight when constrained by a small Lipschitz constant.
%To achieve this Lipschtiz constraint, we use a model of small capacity. Specifically, a neural network of few layers with a small amount of hidden units. In figure~\ref{fig:visualize_weight}, we visualize the weights proposal from our model compared to another competitive baseline. Observe that our model can better captures a semantic tail.

To satisfy the small-Lipschitz constraint, we utilize a model with limited capacity, specifically employing a neural network comprising only a few layers with a few hidden units. Figure~\ref{fig:visualize_weight}B illustrates the weight proposal from our model. The smooth gradient means that the weights are generated under semantic restrictions.
%revealing that our approach accurately produces weights under semantic constraints. %, detailed in Appendix~\ref{sec:visualization},

For robust optimization, we normalize the weights $a(D_{:{i}};\psi)$ as $\bar{\alpha_i}$, taking into account the varying magnitudes of the weights that the adversary might generate:

\begin{equation}
    \bar{\alpha}_i =   \frac{|\gI|a(D_{:i};\psi)}{\sum_{i'} a(D_{:{i'}};\psi)}.
\end{equation}

The semantic weight is important for alleviating the effects of outliers during re-weighting. Across different datasets in our evaluation, we find many outliers possess a low item advantage score. By intuition, these outliers are at disadvantages and we need to increase the weight for those outliers in learning. %However, increasing importance of such outliers during training without similarity between them will greatly reduce the performance of the recommender system. 
However, indiscriminately amplifying the importance of these outliers without accounting for similarities among them can seriously degrade the performance of the recommender system. 
The reason is that the system is not able to generalize to these outliers without adversely affecting the majority. By enforcing a semantically meaningful weight, we promote smoothness and assign similar weights to similar items.  
%In figure~\ref{fig:outliers}, the results show our adversary can focus on the majority of disadvantaged items rather than a handful of outliers. Those outliers are colored by orange and are implicitly identified by the models through semantic constraints.
Figure~\ref{fig:outliers} shows that our adversary is able to concentrate on the majority of disadvantaged items instead of just a few outliers, which are highlighted in green and detected by the models through semantic constraints. As those outliers are distant from the majority, a limited-capacity adversary must choose between promoting the majority of disadvantaged items or focusing on a small number of disadvantaged outliers. As optimization progresses, the strategy of focusing on the majority of items prevails to achieve the goal of maximization.

\begin{figure*}[t]
    \centering
    \begin{subfigure}{0.45\textwidth}
    \includegraphics[width=.95\linewidth]{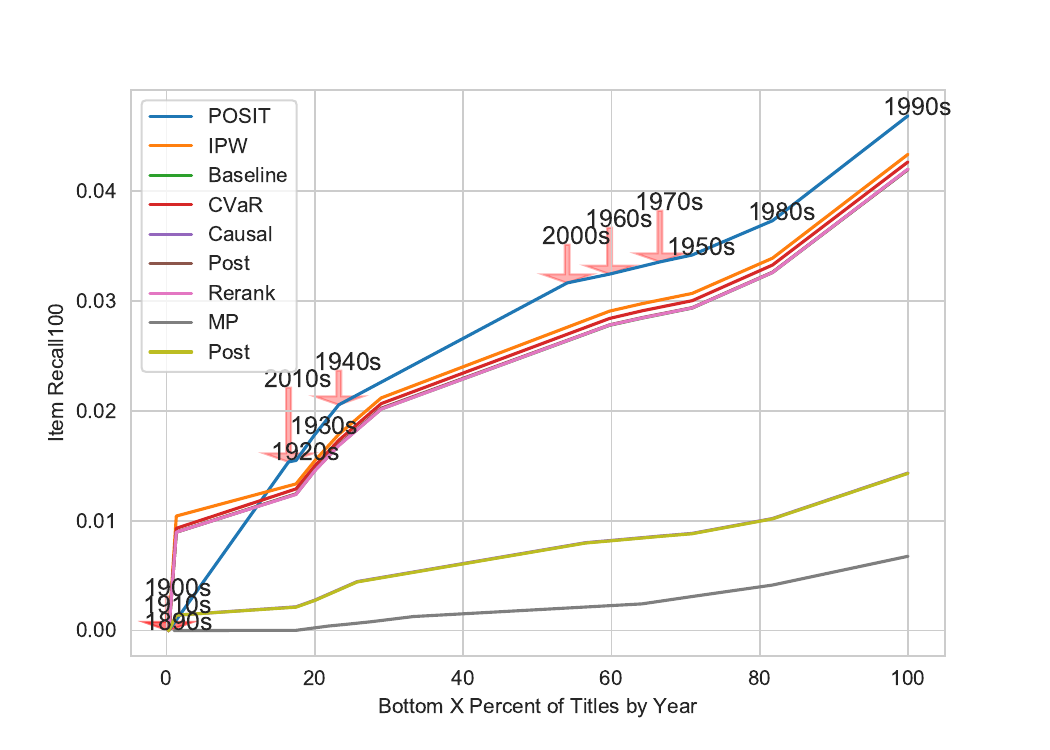}
    \caption{Performance of Different Years}
    \end{subfigure}
    ~
    \begin{subfigure}{0.45\textwidth}
    \includegraphics[width=.95\linewidth]{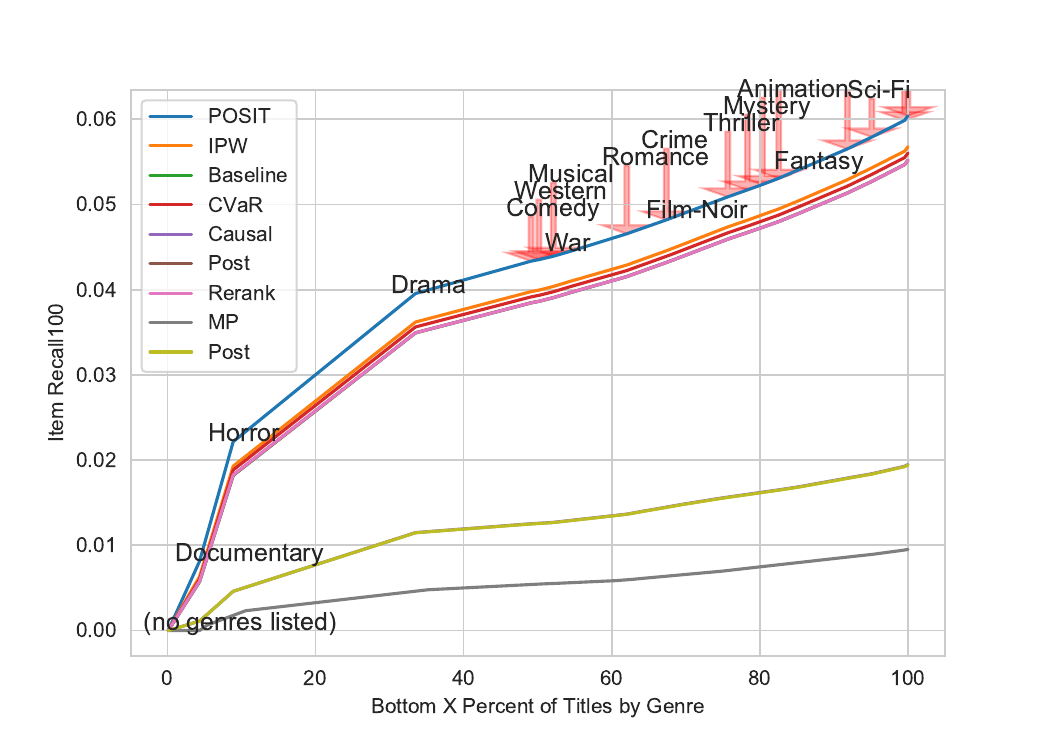}
    \caption{Performance of Different Genres}
    \end{subfigure}    
    \caption{In this figure, we report the Item Recall@100 for movies of different categories. We compare different methods in MovieLens. Performance is averaged for each category and sorted from the worst category to the best. \revision{A low performance on specific categories, such as movies before 1910s and after 2010s, is due to limited data points in the dataset.}}
    \label{fig:category}
    \vspace{-0pt}
\end{figure*}

It should be noted that the advantage score $S_i$ is not a fixed value and is computed against the current model during optimization. In formulation (\ref{eq:general}) we proposed a general form of the learner and adversary that can be specialized to different recommendation base models. In the following, we further rewrite the training paradigm under the baseline EASE by combining eq.(\ref{eq:EASE}) and eq.(\ref{eq:general}):
\begin{align}
    \textrm{Learner}: &\min_{\mW} \sum_{u \in \gU, i \in \gI} \bar{\alpha}_i \, (D_{u:}\mW- D_{u:})_i^2  + \lambda \|\mW\|^2_2  \label{eq:coupling1}  \\
     & \text{s.t.,} \, \textrm{diag}(\mW) = 0 \nonumber \\
    \textrm{Adversary}: & \max_{\psi} \sum_{i \in \gI} \bar{\alpha}_i \,  (-S_i), \quad   \label{eq:coupling2}
\end{align}
where $D_{u:}$ is the user $u$'s interaction history and $\mW$ is the item similarity matrix learned by the EASE model. 
Note that in the summation and $S_i$ in eq.(\ref{eq:coupling1}), we simultaneously optimize for each user and each item. This dependency on both dimensions makes it inappropriate for optimization in batches. %However optimizing the whole dataset is not practical especially for a large dataset. 
To accommodate batch optimization, we further maintain an Exponentially Moving Average (EMA) for values like advantage score $S_i$. These approximate values are gradually updated during iterations.% over batch of users and we will use these approximated values to take place of the true values.

\subsection{Normalization Operation}
 %To further control variation, we further design a normalization operation, which will cap large change to models' outputs and therefore incurs semantic meaningful weights.
%To limit the representation power and stabilize the training, we propose to build a small-capacity adversarial model and combine certain operators that tends to naturally filter out outliers.

Building upon the need to maintain a small Lipschitz constant, we introduce a specific design to further regulate the constant, and thus control the variation in the model's outputs. Our design merges both normalization techniques and bounded activation functions (such as Sigmoid) to naturally filter outliers. This normalization aims to center the input value ranges to zero before activation. Functions such as the sigmoid are employed to bound the output range, capping extremely large inputs, while having minimal impact on values near the average, which often signify outliers.

%Denote the normalization operator as $n(x)=\tau\frac{x-\mu(x)}{\sigma(x)}$ where $\mu$ and $\sigma$ represent mean and standard deviation and $\tau$ an hyper-parameter, Hyperbolic Tangent function as $t(x)$, Sigmoid function as $s(x)$ and Fully Connected Layer as $H(x)$. Such operator combinations can be $t \circ n(x)$ or $s \circ n(x)$ depending on the desired output range. Note that we use an hyper-parameter $\tau$ to further control the degree of deviation of inputs. A larger $\tau$ filter more outliers. For the adversary, we choose $a(x) = t \circ n \circ H_2 \circ s \circ n \circ H_1(x) $, we also compare different choices of structures in Figure~\ref{fig:compare_arch}.

We define the normalization operator as $n(x)=\tau\frac{x-\mu(x)}{\sigma(x)}$, where $\mu$ ($\sigma$) represents the mean (standard deviation) and $\tau$ is a hyperparameter. We also define the hyperbolic tangent function $t(x)$, the sigmoid function $s(x)$, and the fully connected layer $H(x)$. These combinations of operators can be manifested as $t \circ n(x)$ or $s \circ n(x)$, depending on the desired output range. In addition, we utilize the hyperparameter $\tau$ to exert control over the degree of deviation of the inputs. A larger $\tau$ will filter out more outliers. For the adversarial component, we select $a(x) = t \circ n \circ H_2 \circ s \circ n \circ H_1(x)$, and different structural choices are compared in Figure~\ref{fig:compare_arch}.

In the algorithm~\ref{alg:proc}, we show the details of the algorithm. We sample a batch of users in line 4 and estimate the advantage score in line 5. Line 6 normalizes the weights for stable optimization. In lines 7-9, we follow the optimization as described in eq.(\ref{eq:coupling1}). We will update the learning rate and save the best model after each epoch.

\iffalse
Thus using such training loss as an indicator, 

During training, the cross-entropy or mean squared error loss is commonly applied (e.g. in EASE). As a straightforward attempt, similar metric can be reused to tell the advantage, like the mean squared error(MSE) loss for each item. However, we find adopting the same loss function used during training will lead to inadvertent side effects. 

 Thus a high loss value in the training can not faithfully reflect whether this item will be positioned in disadvantage during evaluation. And often we find that such loss value even indicates the opposite way of our expectation.

For example, in Figure xxx, it shows the result of using the mean squared error loss function as the indicator.
\fi

%\ps{IPW, CVar and Rerank may be introduced later after our method. The reason being, after introducing EASE, if we continue with our approach, the flow is good. Otherwise, the reader will be distracted with many other approaches and may bet lost.}
%\ql{TODO: Add description of IPW CVar Rerank after Method.}

%\label{sec:related}
% \usepackage{multirow}
\begin{table}[t]
\begin{tabular}{ccccc}
\toprule
\multirow{2}{*}{Dataset} & \multicolumn{3}{c}{Number of} & \multirow{2}{*}{Interaction Ratio}\\
\cmidrule(lr){2-4}
                         & Users  & Items & Interactions \\
\midrule                    
Movie Lens                & 136k   & 20k   & 10M    &   3.6e-3    \\
Netflix Prize                 & 463k   & 17k   & 57M   &  7.2e-3     \\
Million Song                      & 571k   & 41k   & 34M    & 1.4e-3     \\
\bottomrule
\end{tabular}
\caption{Characteristics of Datasets}
\label{tab:dataset}
\vskip -0.3in
\end{table}

\begin{table*}[t]
\resizebox{.99\textwidth}{!}{% 
\begin{tabular}{cccccccccccc}
\toprule
\multirow{3}{*}{Dataset}       & \multirow{3}{*}{Method} & \multicolumn{10}{c}{Metric}                                                                                      \\ \cmidrule{3-12}
                               &                         & \multicolumn{3}{c}{Coverage@k}                    &  \multicolumn{1}{c}{NDCG@K}   & \multicolumn{3}{c}{Recall@K}  & \multicolumn{3}{c}{Item Recall@k} \\ \cmidrule(lr){3-5}  \cmidrule(lr){6-6}  \cmidrule(lr){7-9}  \cmidrule(lr){10-12}
                               &                         & Top 100       & Top 50        & Top 20    & Top 100  & Top 100     & Top 50        & Top 20  & Top 100     & Top 50        & Top 20         \\
\midrule                             
\multirow{6}{*}{Movie Lens}  & \system          & \textbf{1793} & \textbf{1052} & \textbf{530} &\textbf{0.4214}       & \textbf{0.6369}        & \textbf{0.5244}        & \textbf{0.3928} & \textbf{0.0465}  & \textbf{0.0294}  & \textbf{0.0146} \\
                & IPW           & 1531     & 972      & 519      &\textbf{0.4209}            & \textbf{0.6364}
   & 0.5226             & \textbf{0.3926} & 0.0433  & 0.0270   & 0.0131 \\
                & CVar          & 1488     & 940      & 502      &0.4200            & 0.6353             & 0.5210             & 0.3911 & 0.0426  & 0.0262  & 0.0127 \\
                & Rerank         & 1718     & 948      & 490      &0.4190            & 0.6300             & 0.5199             & 0.3906 & 0.0419  & 0.0257  & 0.0123 \\
                & EASE          & 1461     & 922      & 489      &0.4199            & 0.6356             & 0.5209             & 0.3906 & 0.0419  & 0.0257  & 0.0123 \\
               & Causal   & 1084      & 417  & 259   & 0.3695    & 0.5599     &      0.4560   &   0.3411   & 0.0143 &  0.0083 & 0.0041   \\
               & Post   &        1586 &  922 & 489 & 0.3927 & 0.5525 & 0.4448 & 0.4118 & 0.0419 & 0.0257 & 0.0123  \\
                & MP     & 100     &  50    &  20  &0.1905       &    0.3300        &  0.2351           &      0.1617 & 0.0067  & 0.0035  & 0.0016 \\
\cmidrule(lr){1-12}                
\multicolumn{2}{c}{Standard Deviation}                 & $\pm 10$   & $\pm 5$    & $\pm 2$    &\multicolumn{4}{c}{$\pm 0.0009$} & \multicolumn{3}{c}{$\pm 0.0004$} \\
\midrule
\multirow{6}{*}{Netflix Prize} & \system          & \textbf{1968} & \textbf{1256} & \textbf{694} &\textbf{0.3953}       & \textbf{0.5565}        & \textbf{0.4472}        & \textbf{0.3633} & \textbf{0.0958}  & \textbf{0.0631}  & \textbf{0.0342} \\
                & IPW           & 1802     & 1179     & 670      &0.3942            & \textbf{0.5560}             & \textbf{0.4464}             & \textbf{0.3625} & 0.0833  & 0.0526  & 0.0269 \\
                & CVar          & 1795     & 1177     & 667      &0.3933            & 0.5537             & 0.4449             & 0.3617 & 0.0842  & 0.0532  & 0.0271 \\
                & Rerank         & 1950     & 1190     & 655      &0.3919            & 0.5478             & 0.4435             & 0.3612 & 0.0802  & 0.0502  & 0.0252 \\
                & EASE          & 1762     & 1153     & 654      &0.3930            & 0.5541             & 0.4448             & 0.3613 & 0.0804  & 0.0502  & 0.0252 \\
               & Causal & 1394 & 962 & 564   & 0.3551 & 0.5034 & 0.4024 & 0.3233 & 0.0251 & 0.0468 & 0.0249\\
               & Post & 1759 & 1155 & 654 & 0.3927 & 0.5525 & 0.4448 & 0.3612 & 0.0777 & 0.0468 & 0.0249\\
                & MP     & 100     &  50    &  20           &0.1587       &    0.2743        &  0.1749           &      0.1161 & 0.0083  & 0.0044  & 0.0019 \\
\cmidrule(lr){1-12}                 
\multicolumn{2}{c}{Standard Deviation}                 & $\pm 6$    & $\pm 4$    & $\pm 2$     &\multicolumn{4}{c}{$\pm 0.0009$} & \multicolumn{3}{c}{$\pm 0.0004$} \\
\midrule
\multirow{6}{*}{Million Song} & \system          & \textbf{6569} & \textbf{3692} & \textbf{1625}     &\textbf{0.3915}      & \textbf{0.5116}      & \textbf{0.4290}      & \textbf{0.3341} & \textbf{0.2479}   & \textbf{0.2030}   & \textbf{0.1433} \\
                & IPW           & 6466     & 3673     & \textbf{1626} &\textbf{0.3909}           & 0.5099           & \textbf{0.4290}          & \textbf{0.3340} & 0.2464   & 0.2018   & 0.1419 \\
                & CVar          & 6337     & 3556     & 1567     &0.3839           & 0.5070           & 0.4240
& 0.3268 & 0.2401  & 0.1917   & 0.1314 \\
                & Rerank         & 6368     & 3624     & 1609     &0.3896           & 0.5062           & 0.4278
 & \textbf{0.3339} & 0.2444   & 0.2007   & 0.1412 \\
                & EASE          & 6304     & 3623     & 1609   &0.3898           & 0.5084           & 0.4279          & \textbf{0.3339} & 0.2445   & 0.2007   & 0.1412 \\
               & Causal & \multicolumn{10}{c}{-} \\
               & Post & 6342 & 3632 & 1611 & 0.3899 & 0.5072 & 0.4279 & 0.3340 & 0.2453 & 0.2013 & 0.1416 \\                
                & MP     & 100     &  50    &  20                     &0.0582       &    0.0986        &  0.0680           &      0.0427 & 0.0027 & 0.0018  & 0.0010 \\
\cmidrule(lr){1-12}                 
\multicolumn{2}{c}{Standard Deviation}                 & $\pm 9$    & $\pm 5$    & $\pm 2$    &\multicolumn{4}{c}{$\pm 0.0009$} &  \multicolumn{3}{c}{$\pm 0.0004$} \\
\bottomrule
\end{tabular}
}
\caption{We report four metrics, including coverage and both user-side and item-side performance, for five different methods in this table. Here, $\pm$ represents the standard deviation of the metric. To calculate the standard deviation of Coverage, we split the test set in chunks of 100 users. We report the standard deviation between chunks. For performance, we run each baseline 10 times on different test sets. Best results within one standard deviation are highlighted. We found NDCG@\{20,50\} reflects similar trends as NDCG@100. Please refer to Appendix due to space limit.  Causal fails on Million Song due to memory limit.}
\label{tab:total}
\vskip -0.20in
\end{table*}

\section{Baselines}
\label{sec:baselines}

In this section, we introduce and compare against six other baselines IPW~\citep{IPW}, CVar~\citep{rockafellar2000optimization}, Rerank~\citep{adomavicius2009toward}, Causal~\citep{causal2}, Post~\citep{post} and MP.

\textbf{IPW}. Inverse Propensity Weighting (IPW)~\citep{IPW} assigns a weight inversely proportional to the interaction frequency of an item. Thus, infrequent items receive a larger weight. To make this method more competitive, we further apply a power $\beta$ to the number and a normalization. We search and report the best $\beta$. Formally, the probability of interactions for an item $j$ can be counted as $\frac{1}{|\gU|}\sum_{i\in\gI}D_{ij}$. The following weight $\bar{w}_j$ is applied to each item $j$ during training:
\begin{equation}
    w_i = \left(\frac{1}{|\gU|}\sum_{u\in\gU}D_{ui}\right)^{\beta},  \quad
    \bar{w}_i = |\gI| w_i \mathbin{/} (\sum_{i \in \gI} w_i) \, .
\end{equation}

\textbf{CVaR}. The condition value at risk (CVaR)~\citep{rockafellar2000optimization} optimizes over only the tail of items. A tail contains the worst $\alpha$ percent performed items, which are measured by the training loss function.  Formally, consider the loss function $\sum_u \gL(D_u;\mW)_i$ as in eq.(\ref{eq:general_loss}), a positive indicator $[x]_+ = \max(0,x)$  and an additional variable $\beta_1$, we have the optimization goal
\begin{equation}
\min_{\mW,\beta_1} \sum_{i\in \gI}\left\{\beta_1 + \frac{1}{\alpha}\left[\sum_{u \in \gU} \gL(D_u;\mW)_i - \beta_1\right]_+\right\} \, .
\end{equation}
An optimal solution will ensure that $\beta_1$ is the threshold of training loss which ensures that only the tail part of the loss is optimized.
%Mathematically, we have
%\begin{equation}
%    \Pr\left(j\in \gI | \sum_{i \in \gU} %%\gL(D_i;\mW)_j \geq \beta_1\right) = \alpha \, .% \nonumber
%\end{equation}

\textbf{Rerank}. Rerank~\citep{adomavicius2009toward}   promote items in the tail by post-processing during evaluation. Suppose that a recommender ranks all items into a list for some user based on relevance $[i_1, i_2, ..., i_n]$. Each item is associated with a relevance score $s$ such that $s_1 \geq s_2 ... \geq s_n$.  It partitions this recommendation results by two thresholds $T_h$ and $T_l$. The top recommendations are items of relevance $s_i \geq T_h$. The method will keep their order the same as before. The tail recommendations are items of relevance $T_h \ge s_i \geq T_l$. They are resorted according to frequency of the item's occurrence. Infrequent items will rank higher than before, but still rank lower than the top recommendations. Other recommendations with even less relevance are simply ignored, as they represent irrelevant items. We fine-tune the two thresholds during evaluation and report the best result.

\textbf{Causal}. Causal elimination~\citep{causal2} mitigates popularity bias by employing a causal graph.  Within this graph, popularity is introduced as an intricate interaction that includes users, items, and outcomes. They further estimate the causal effect through history and mitigate the popularity effect by canceling part of the impacts.

\textbf{Post}. Post-processing~\citep{post} modifies the final recommendation scores by introducing a popularity neutral bias. This bias is derived from the original recommendation score, with the intention of shifting in favor of less popular items.

\textbf{MP}. The Most-Popularity (MP) algorithm ranks items solely based on their popularity, with more popular items receiving a higher rank. This baseline is included to provide an intuitive demonstration of the metrics on vanilla baselines.

\section{Experiments}
\label{sec:expts}

In this section, we report experiments on three large-scale datasets (summarized in Table~\ref{tab:dataset}), namely, MovieLens-20M~\citep{movielens}, Netflix Prize~\citep{netflixprize} and the Million Song Dataset~\citep{millionsong} in the baselines introduced in Section \ref{sec:baselines}. In the comparison, we conduct parameter sweeps for each baseline and report test metrics results for the best model on validation set. We follow the same procedures for all baselines except CVaR and Rerank. They have direct trade-offs between performance and coverage. Reporting the model with the best performance will only regress to the baseline EASE. Instead, we report a representative model with comparative performance. %\ps{QL: how is good balance defined?}\ql{Changed to find a model with comparative performance.} \ps{QL: was the best model picked such that it had the best overall metrics (NDCG or recall) for each method?}. \ql{No, because rerank and CVar is a direct trade-off. Doing so will however fallback to the same baseline} The characteristics of each dataset are summarized in Table~\ref{tab:dataset}. During training, we did early stops when no metric improvement are made after 10 training epochs on the validation set.  

Specifically, we initialize the parameters $\mW$ of EASE and the first layer of the adversary model as $0$. The reason is that these parameters directly take sparse inputs.  Randomly initialized parameters will prevent a stable convergence, as some of these values may hardly get a chance to update themselves. We use SGD optimizer for our approach. For all baselines, we fine-tune the hyperparameters, including the learning rate and $\ell_2$-regularized $\lambda$. For \system, we also tuned the learning rate of the adversarial model and the structure of the model (see Figure~\ref{fig:compare_arch}). For IPW, we tune the parameter $\beta$. We tune the parameter $\alpha$ for CVaR and $T_h$ and $T_l$ for Rerank. For Causal, we tune $\alpha$ and $\beta$. For Post, we tune $\beta$ and $\gamma$. The subpar performance of the Causal method is likely attributed to its reliance on the strong assumption that the model's output carries a probabilistic meaning, an assumption that is incorrect under EASE's framework. Additional details can be found in ~\autoref{app:exp_details}. %A~\footnote{See \url{https://anonymous.4open.science/r/data-DF78/Appendix.pdf}}.%

%More details of experiments can be found in Appendix~\ref{app:exp_details}. 

%\ps{QL: can you add more details on the experiments? How were the models trained? How were the hyper-parameters tuned for each dataset/method? What was the stopping criterion for the ADVISE method and may be for other methdos etc. How were the parameters intialized? The idea is that we provide as much information as possible towards reproducibility.} \ql{Sure, I will add. Should we add the details in appendix? Since there are a lot more details.}
%We can observe  xxx based on\ql{ToDO: Verify the information in dataset and conclude how it relates to the result.} .

%\ref{tab:coverage}
\textbf{Coverage.}
In Table~\ref{tab:total} we report the coverage@k from different methods on the three datasets. We report coverage@k when $k=20,50,100$ for a thorough evaluation. The calculation of coverage can be found in eq.(\ref{eq:coverage_def}). The result shows that our proposed method consistently achieves a competitive coverage number compared to other baselines on three datasets. Specifically, our method improves nearly 20\% on Coverage@100 compared to that of EASE on MovieLens dataset.While Rerank has a coverage that is close to POSIT when $k=100$, its trade-off towards coverage significantly impacts its performance.
By comparing the results, we observe that the proposed method is much more effective in scenarios where k is large or when the baseline has poor coverage. For example, the Coverage@100 of EASE on MovieLens is one-quarter of that number on Million Song dataset. Consequently, the increase in coverage@100 in MovieLens is greater than on Million Song. The benefits of promoting tails increase as the baseline becomes more concentrated due to the training data.

Similarly, the improvement of Coverage@100 is greater than that of Coverage@20 on these three datasets. It is because the recommender rarely ranks the items from the tail to the very top places. Therefore, the benefits of promoting items from the tail will gradually add up to a larger $k$ value for some metric@k. The metric with larger $k$ is quite important for users who explore deep into the catalog, such as frequent users or users with many interests.

\iffalse
\begin{figure*}[t]
    \centering
    \begin{subfigure}{0.32\textwidth}
    \includegraphics[width=.95\linewidth]{figures/ml20m_compare_metric.png}
    \caption{Comparison of Advantage Score}
    \label{fig:compare_adv_score}
    \end{subfigure}
    ~
    \begin{subfigure}{0.32\textwidth}
    \includegraphics[width=.95\linewidth]{figures/ml20m_compare_model.png}
    \caption{Comparison of Model Architecture}
    \label{fig:compare_arch}
    \end{subfigure}
    ~
    \begin{subfigure}{0.32\textwidth}
    \includegraphics[width=.95\linewidth]{figures/ml20m_compare_units.png}
    \caption{Comparison of Model Capacity}
    \label{fig:compare_capacity}
    \end{subfigure}
\end{figure*}
\fi 
%\ref{tab:performance}
\textbf{Performance.}
Note that it is commonly assumed there is a trade-off between coverage and accuracy metric Recall@k. However, we find that \system achieves the best of both worlds. Table~\ref{tab:total} reports the performance metrics, including Recall@K and NDCG@k.   Observe that our method achieves slightly better performance than all the other baselines on all three datasets. We observe a similar trend as in the case of coverage, where the improvement is larger given a large $k$. The improvement of \system is statistically significant (two standard deviation) on Netflix Prize, but not necessarily on the other two datasets. The reason is that popular items consist of the majority of interactions. Consequently, although our method has greatly improved the tail, the improvement is not necessarily reflected in the metric after averaging over the whole dataset.
%Also note that we mainly use Recall@k since NDCG is strongly correlated with Recall. 
To overcome this issue, we further report the Item Recall@k (see eq.(\ref{eq:itemrecall})).%which focuses on the tail in Table~\ref{tab:total}.%~\ref{tab:perf_item}.  
Unlike the Recall, this metric reflects the performance of the recommender on the tail. This is a more suitable metric, as all baselines aim to improve the tail. The result shows that our method greatly outperforms the other baselines. Specifically, we improve performance at most 24\% compared to EASE, while the other methods do not have a comparable result.
%\ps{QL: by excluding popularity bias do you mean using equation \eqref{eq:advantage_score_wo_popularity}, please clarify and refer to the equation if that is the case.} \ql{Yes}

\begin{figure}[t]
\vskip -0.15in
\includegraphics[width=.83\linewidth]{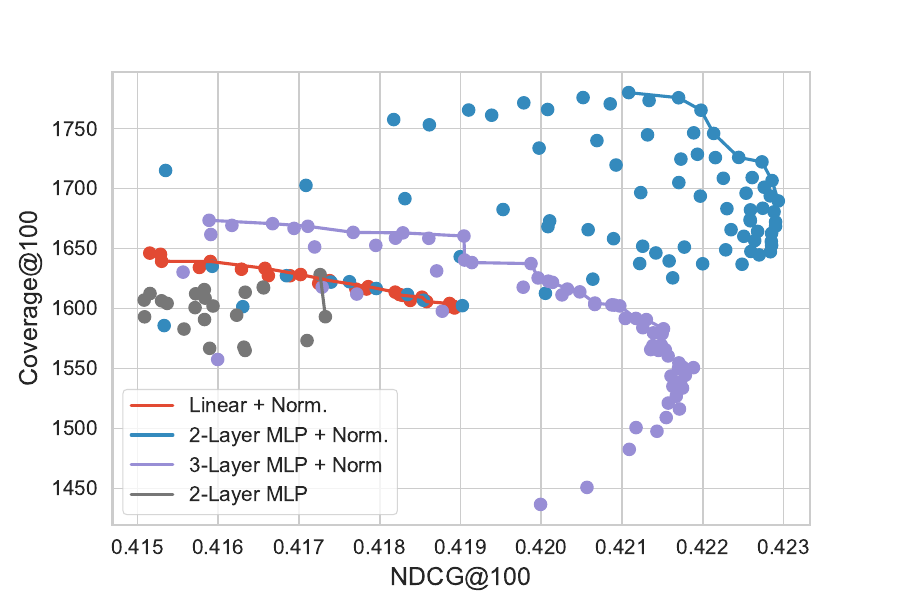}
\vskip -0.15in
\caption{Comparison of Model Architecture. "Norm" indicates an $\ell_2$ standardization applied before nonlinearity activation. "MLP" indicates multi-layer perceptron. We use the tan-hyperbolic activation for intermediate layers and Sigmoid for the final output.}
\label{fig:compare_arch}
\vskip -0.15in
\end{figure}

\textbf{Semantic Group.} In Figure~\ref{fig:category}, we report performance in different categories of items on MovieLens. This dataset provides additional attributes associated with each movie, including released years and genres. Since they constitute some kind of semantic information, we look at the performance over titles grouped by different years/genres.  It is worth noting that semantic information such as year and genre is not used by \system during learning. The result shows that semantically meaningful weights help to improve the performance of different semantic groups without explicit labels, whereas the baseline methods do not show any improvement for most of these groups. In Appendix~\ref{sec:freq}, we also report the Gini index. We find that our model increases the fairness of the prediction.
%\ps{provide more details of this result. May be say that in one of the datasets we have year and genre associated with each title.}\ql{Changed}
\begin{figure}[t]
\vskip -0.15in
\includegraphics[width=.83\linewidth]{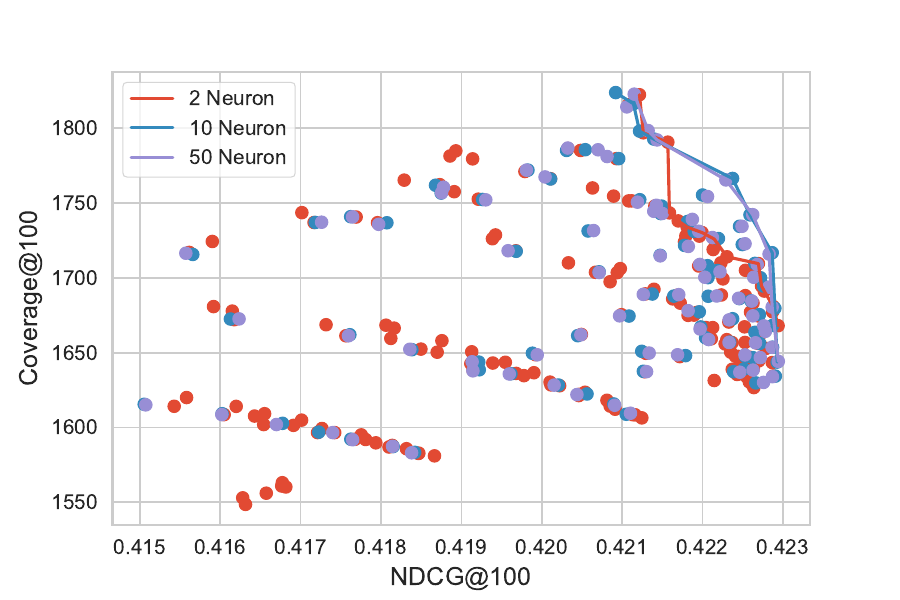}
\vskip -0.15in
\caption{Comparison of Model Capacity. The label shows the number of hidden units on a 2-layer fully connected neural network for the adversary.}
\label{fig:compare_capacity}
\vskip -0.17in
\end{figure}

\subsection{Ablation Study}
\label{sec:ablation}
In this section, we conduct ablation studies on various %
research questions(RQ) impacting the \system. We study one factor at a time, perform parameter sweeps for other factors, and report the best result. We will draw the Pareto frontier~\cite{pareto} between coverage and performance. A curve in the upper right corner is better.

\textbf{RQ1: How to design a good adversarial model for \system?}

We evaluated different adversarial model structures in Figure~\ref{fig:compare_arch} and found that combining bounded activation and normalization leads to improved performance. This suggests that these operators effectively filter out outliers. Observe the large differences in coverage between the 2-layer models with or without such operators in Figure~\ref{fig:compare_arch}. It is worth noting that the seemingly small variation on NDCG has a big impact on semantic tails, due to their minor presence. It also showed that the number of layers, or model capacity, is a crucial factor for the best performance. Specifically, a two-layer MLP provides the best results.
We also find that a small model (with 10 hidden units) provides good performance in Figure~\ref{fig:compare_capacity}. As the size of adversary is around $1\textrm{e-}3$ of the size of the recommender, the time overhead of the adversary is small.

%We compare different structures of the adversarial model in Figure~\ref{fig:compare_arch}. The results in Figure~\ref{fig:compare_arch} show that the combination of bounded activation and normalization improves the performance, supporting the hypothesis that these operators can better filter out outliers. Comparing adversarial models of different layers, we find that the model capacity, denoted as the number of layers, is a crucial factor for the best performance. The result shows that a model capacity of 2-layer MLP gives the best result.
%We further find that a small model (with 10 hidden units) provides a good performance in Figure~\ref{fig:compare_capacity}. As the size of adversary is around $1\textrm{e-}3$ of the size of the recommender, the time overhead of adversary is small. 

Additional studies of the research question can be found in the \autoref{sec:more_ablation};  The second research question (RQ2) investigates whether popularity bias is essential to identify a disadvantaged item. Our findings show that even \textit{ without specific information about popularity}, an adversary can recognize less popular items as disadvantaged. This suggests potential common interaction patterns for disadvantaged items other than just popularity. The third research question (RQ3) explores the impact of the batch size of the user on the definition of coverage and its effect on the results presented. Our analysis reveals a trade-off between inter-user diversity and the coverage of rare items, depending on the number of users selected in the batch. Furthermore, Appendix \ref{sec:visualization} compares the use of an adversary model with the advantage scores. The results demonstrate a notable improvement over smooth weight assignment through the adoption of the adversary model in visualization.

\section{Conclusions}
\label{sec:conclude}
We proposed an adversarial learning approach called POSIT to reduce popularity bias by identifying semantic items that form tails. A key challenge was to take a user-level metric and convert it to an item-level score that could be exploited in an adversarial training. We analyze how our method works through visualization and studying different aspects of the problem. POSIT was shown to achieve significantly better coverage and even improved overall performance in three publicly available large-scale datasets.

%\title{Appendix}

\clearpage
\bibliographystyle{plain}
\bibliography{sample-base.bib}

\clearpage

\appendix

%\ps{Do they want one column for the Appendix?}
%\ql{They don't say the format of appendix. It is required to upload at the other places independently. Two column is kind of ugly.}
\section{Experiment Details}
\label{app:exp_details}
%\ps{Let us give really detailed information here for reproducibility. Also, hope you can make the code public assuming the paper gets accepted.}
In this section, we discuss the experiments configuration in details.
We set training epochs $T=50$, $m=0.9$ in algorithm~{\ref{alg:proc}}. We use a batch size of 1024 for MovieLens and 8192 for Million Song and Netflix Prize. We set the momentum of SGD optimizer as 0.9. We use 10k users for validation and another 10k users for testing on Movie Lens. We increase this number 10k to 40k on Netflix Prize and 50k on Million Song dataset, respectively. Some datasets provide explicit preference information, like Movie Lens. We follow the literature~\cite{EASE} and convert them into implicit settings using a preference threshold. Preference score greater than this threshold is set to 1 and otherwise it is set to 0.

We conducted the experiments on machines with 64G memory and 64 CPUs without GPU. This choice is due to budget concerns. We can finish the training on CPU as our model is relatively small and reasonably fast enough on CPU. It typically takes 8 hours to train \system on MovieLens, 1.5 days on Netflix Prize, and 1 day on Million Song. The major overhead is due to gradient optimization on these very large datasets. Speed can be potentially improved vastly with GPU based training.

\begin{figure}[h]
\includegraphics[width=.82\linewidth]{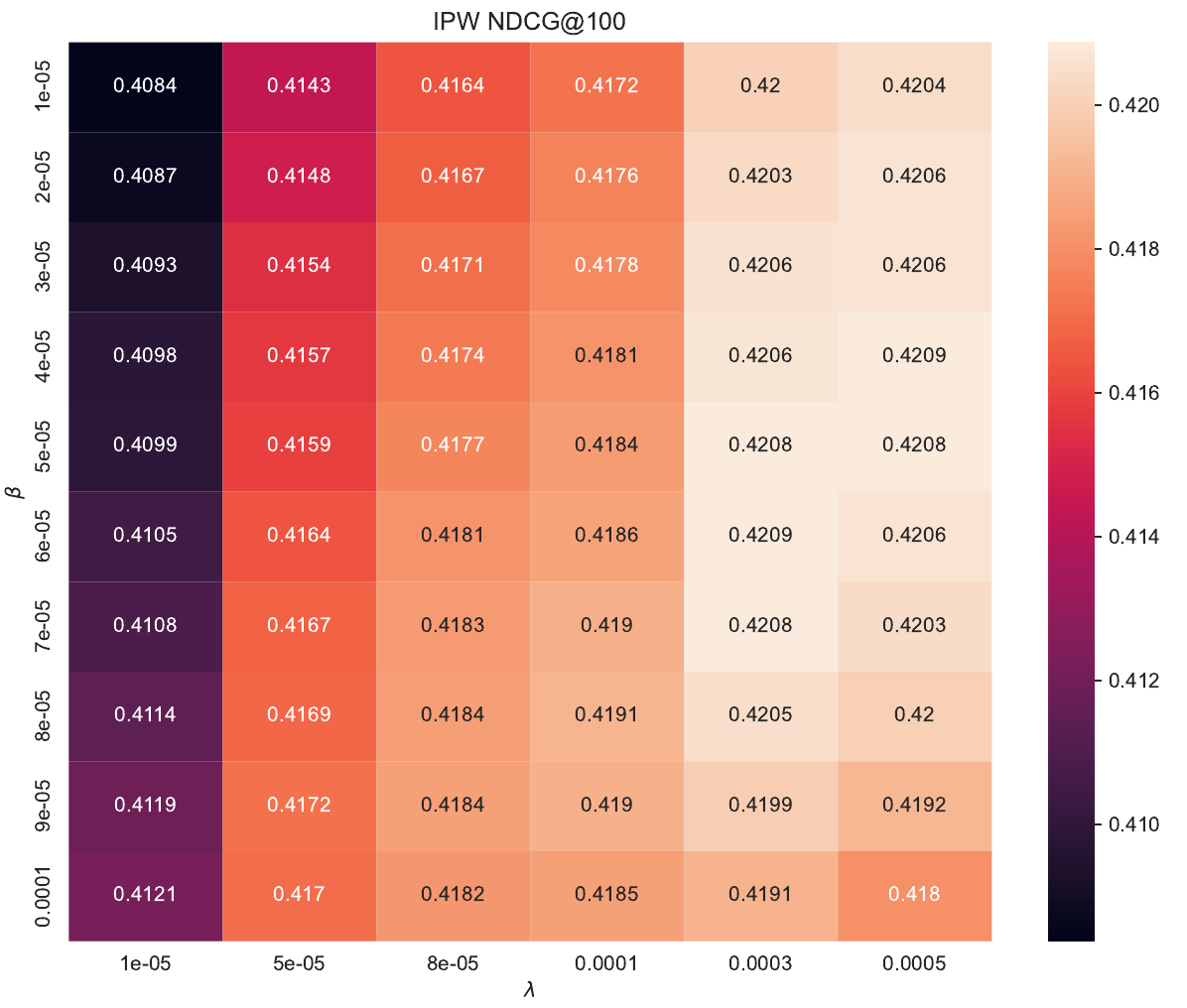}
\caption{Example of Hyper-parameter Search for IPW on MovieLens}
\label{fig:param_search_ml20m}
\vskip -0.05in
\end{figure}

We conduct hyperparameter searches. Figure~\ref{fig:param_search_ml20m} is an example of the parameter search for baseline IPW on MovieLens. Note that we may fine-tune multiple factors and the results will not necessarily be presented as a 2-dimensional figure. Table~\ref{tab:hyper_param} lists all the hyperparameters that are needed to exactly reproduce our results. We will also release our codes upon acceptance.

%Figures~\ref{fig:param_search_ml20m} and \ref{fig:param_search_nflx} are two examples of the parameter search for baseline IPW on MovieLens and NFLX Prize. Note that the best parameter can be found around the center of the figure. And we can empirically observe changing hyper-parameter in any other directions will simply bring down the performance. Note that we may fine-tune multiple factors and the results won't necessarily be presented as a 2-dimensional figure. For example, for baseline IPW, we also fine-tune the learning rate. Therefore, we empirically conduct hyper-parameter searching for multiple times to better pinpoint an suitable range on different hyper-parameter combinations.

\iffalse

\begin{figure}[h]
\includegraphics[width=.95\linewidth]{figures/IPW_nflx_ndcg.pdf}
\caption{an Example of Hyper-parameter Search for IPW on NFLX Prize}
\label{fig:param_search_nflx}
\end{figure}
\fi

\begin{table}[h]
\begin{tabular}{cccc}
\toprule
                          & Movie Lens              & Netflix Prize           & Million Song \\
\midrule
Learning Rate(LR)             & 2.0                     & 1.0                     & 40.0         \\
Adversarial LR & 1.0                     & 1.0                     & 1.0          \\
$\lambda$                    & 8e-6                    & 5e-5                    & 1e-5         \\
Advantage Score           & As in Eq.(\ref{eq:advantage_score_with_popularity}) & As in Eq.(\ref{eq:advantage_score_with_popularity}) &      As in Eq.(\ref{eq:advantage_score_with_popularity})        \\
$\tau$                       & 1.5                     & 1.0                     & 0.5          \\
Adversarial Model         & 2 Layer MLP    &      2 Layer MLP                   &     2 Layer MLP \\
\bottomrule
\end{tabular}
\caption{Parameters used in our evaluation. }
\label{tab:hyper_param}
\end{table}

\section{Frequency Diversity}
\label{sec:freq}
In table~\ref{tab:gini}, we report the relative Gini Index of recommenders~\citep{gini_index,mcauley2022}. It is the Gini Index on recommendations divided by the same index on training data. It represents the increase in popularity-bias for a recommendation system. A smaller value is better. A value greater than 1 indicates the learning algorithm increases the popularity bias.
We can observe that our algorithm tends to have a smaller number compared to other baselines. %\ps{Incomplete sentence.}
%\ps{what do you mean by that implied from the original dataset?}.\ql{We can calculate the Gini index on frequency of training dataset. It shows how fairness issue exists on the training data}

%\newpage

% \usepackage{multirow}
\iffalse
\begin{table}[t]
\begin{tabular}{ccccccc}
\multirow{2}{*}{Dataset}       & \multirow{2}{*}{Metric} & \multicolumn{5}{c}{Method}                      \\
                               &                         & Ours           & IPW   & CVaR  & Rerank & EASE  \\
\multirow{2}{*}{Movie Lens}    & Gini Ratio              & \textbf{1.033} & 1.041 & 1.042 & 1.039  & 1.043 \\
                               & Std.      & \multicolumn{5}{c}{$\pm 0.001$}                 \\
\multirow{2}{*}{Netflix Prize} & Gini Ratio              & \textbf{1.077} & 1.087 & 1.089 & 1.088  & 1.088 \\
                               & Std.      & \multicolumn{5}{c}{$\pm 0.002$}                 \\
\multirow{2}{*}{Million Songs} & Gini Ratio              & \textbf{0.941} & 0.958 & 0.995 & 0.971  & 0.977 \\
                               & Std.      & \multicolumn{5}{c}{$\pm 0.002$}                
\end{tabular}
\end{table}
\fi

% \usepackage{multirow}
\begin{table}[h]
\begin{tabular}{cccc}
\toprule
\multirow{2}{*}{Method} & \multicolumn{3}{c}{Gini Ratio}                   \\
\cmidrule(lr){2-4}
                        & Movie Lens     & Netflix Prize  & Million Songs  \\
\midrule
\system                   & \textbf{1.033} & \textbf{1.077} & \textbf{0.941} \\
IPW                     & 1.041          & 1.087          & 0.958          \\
CVaR                    & 1.042          & 1.089          & 0.995          \\
Rerank                  & 1.039          & 1.088          & 0.971          \\
EASE                    & 1.043          & 1.088          & 0.977          \\
Causal                  & 1.053          & 1.104          & -              \\
Post   & 1.039  & 1.089 & 0.969 \\
MP                      & 1.089          & 1.146          & 1.662          \\
\cmidrule(lr){1-4}
Standard Deviation      & $\pm 0.001$    & $\pm 0.002$    & $\pm 0.002$   \\
\bottomrule
\end{tabular}
\caption{In this table, we show the Gini ratio between original dataset and the prediction from the model. 
}
\label{tab:gini}
\vskip -0.3in
\end{table}

\section{Additional Ablation Studies}
\label{sec:more_ablation}
In this section, we include more ablation studies here.

\begin{figure}[h]
\vskip -0.10in
\includegraphics[width=.95\linewidth]{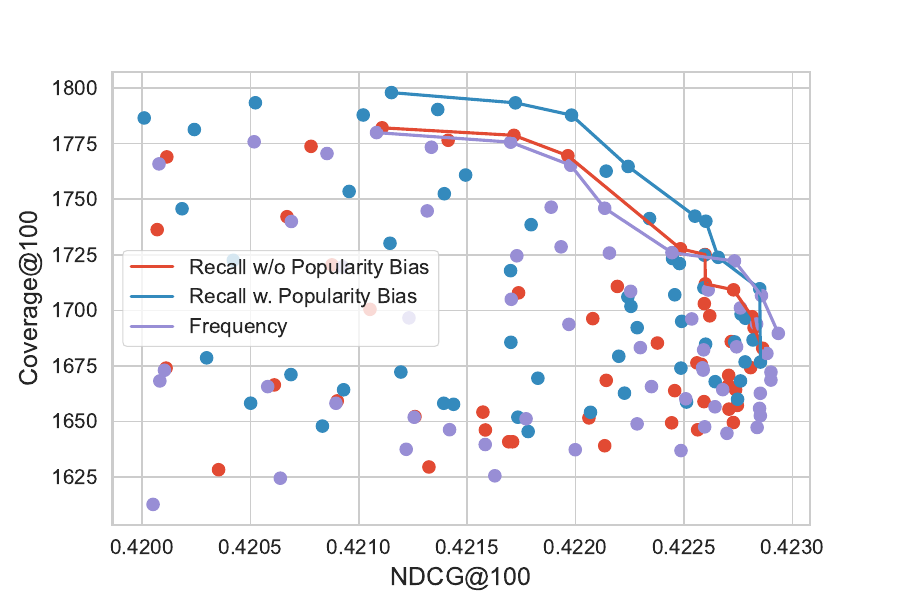}
\vskip -0.20in
\caption{Comparison of Advantage Score}
\label{fig:compare_adv_score}
%\vskip -0.05in
\end{figure}

\begin{figure*}[t!]
\begin{subfigure}{0.3\textwidth}
\includegraphics[width=.95\linewidth]{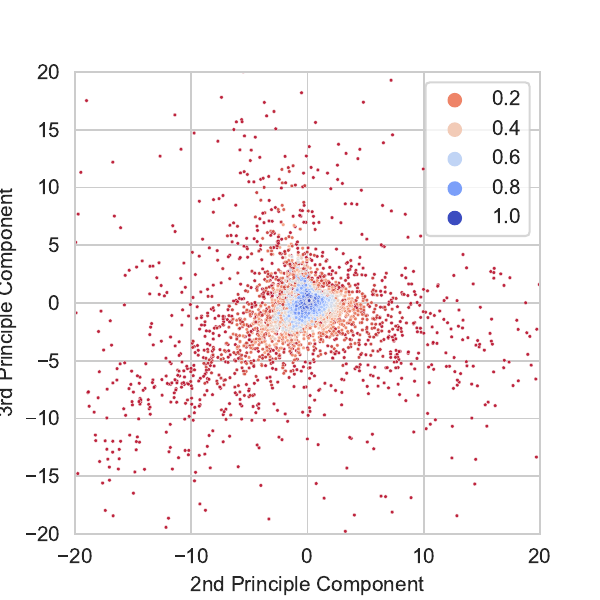}
\caption{Weight from \system}
\end{subfigure}
~
\begin{subfigure}{0.3\textwidth}
\includegraphics[width=.95\linewidth]{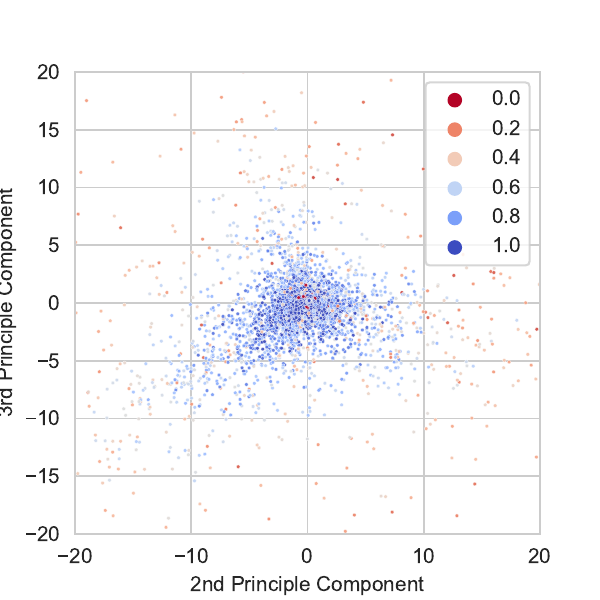}
\caption{Normalized Negative Advantage Score}
\end{subfigure}
~
\begin{subfigure}{0.3\textwidth}
\includegraphics[width=.95\linewidth]{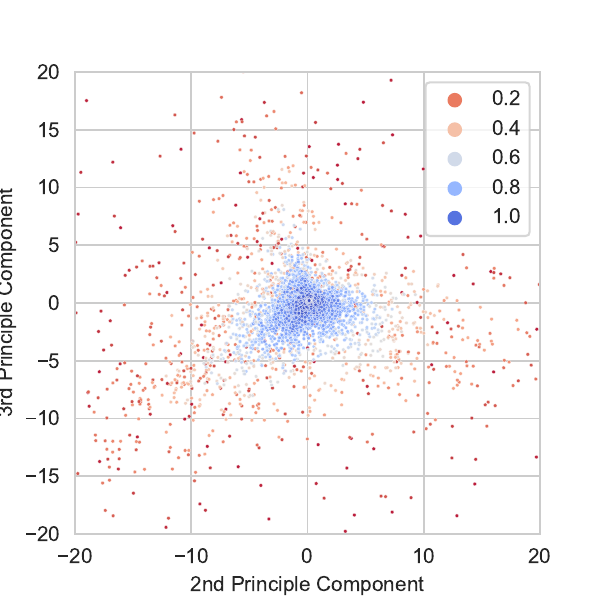}
\caption{Weight from the Best IPW model}
\end{subfigure}
\caption{Semantically Meaningful Weight. In this figure, we visualize the movies and their corresponding weights in the MovieLens-20M dataset obtained from different baselines.  We visualize the movies by conducting Primary Component Analysis (PCA) on features of the movies. Each movie is represented as a point. We use the second and third components as the x-axis and y-axis. Note that the first component is omitted because it is less related to semantic similarity. We discuss this issue in detail in ~\autoref{fig:visualization_first_second}.  }
\label{fig:visualize_weight_comparison}
\vskip -0.1in
\end{figure*}

\textbf{RQ2: Does the adversarial model require popularity bias to determine the weight?}

We compare three candidates of the advantage score on the same model in Figure~\ref{fig:compare_adv_score}.  After hyperparameter tuning, the results show that the advantage score without popularity bias performs similarly to the frequency score. Note that the frequency score is a key indicator of popularity bias. This supports the idea that the score derived from user-level performance without popularity bias can effectively separate advantaged items from the disadvantaged ones. Furthermore, by combining popularity bias in eq.(\ref{eq:advantage_score_with_popularity}), we achieve better results than relying solely on popularity bias.

\revision{\textbf{RQ3: How will the batch size ($|\gU|$) impact the results of Coverage@k  in its definition (eq.~\ref{eq:coverage_def})?}} \\
The batch size ($|\gU|$ in (eq.~\ref{eq:coverage_def})) used to calculate Coverage@k is fixed at 100 for fair comparisons. However, different batch sizes may emphasize different aspects of coverage. When the batch size is small, such as 2, the metric focuses more on inter-user diversity. As the batch size increases larger, such as 100, the metric focuses on coverage of rare items. To gain a better understanding of \system's induction bias, we compare the Coverage@k under different user batch sizes, as shown in Table~\ref{tab:coverage_batch_size}. The results show that \system consistently produces competitive results in different batch sizes. This suggests that using batch size 100 strikes a good balance between dissimilarity and rare-item coverage. Furthermore, we observe an improvement 60\% on \system compared to EASE when the batch size is 100. This highlights \system's ability to better promote rare items.
%In the definition of Coverage@k, We fix the batch size $|\gU|$ as 100 for fair comparisons. However, different batch sizes may stress different perspectives of the coverage. When the batch size equals to 2, this metric represents expected dissimilarity between two users' recommendations. As the batch size becomes larger, this metric shifts to the coverage of rare items. To better analyze the advantage of \system, we compare different user batch sizes and report the results in Table~\ref{tab:coverage_batch_size}. Interestingly, we find that \system consistently has competitive results under all batch sizes. This suggests using batch size 100 provides a good balance towards both ends. Furthermore, \system has larger improvements given a larger batch size. This suggests that \system can better promotes those rare items.  

\begin{table}[h]
\begin{tabular}{ccccccc}
\toprule
\multirow{2}{*}{Method} & \multicolumn{5}{c}{User Batch Size in Coverage@100}                   \\
\cmidrule(lr){2-6}
                        & 10     & 50  & 100 & 500 & 1000  \\
\midrule
\system                   & \textbf{560} & \textbf{1309} & \textbf{1793} & \textbf{3386} & \textbf{4245}\\
IPW                     & 549          & 1190          & 1531   &     2431  & 2863      \\
CVaR                    & 543          & 1163          & 1488   &      2330 & 2734      \\
Rerank                  & \textbf{562}          & 1284          & 1718   &   3108      & 3896   \\
EASE                    & 540          & 1147          & 1462   &   2260 &2635        \\
\cmidrule(lr){1-6}
Standard Deviation      & $\pm 2$    & $\pm 5$    & $\pm 10$ &  $\pm 18$ & $\pm 40$  \\
\bottomrule
\end{tabular}
\caption{In this table, we evaluate the Coverage@100 by varying user batch sizes on the MovieLens-20M dataset.}
\label{tab:coverage_batch_size}
\vskip -0.3in
\end{table}

\section{Visualization}
\label{sec:visualization}

\begin{figure}[h]
\includegraphics[width=.75\linewidth]{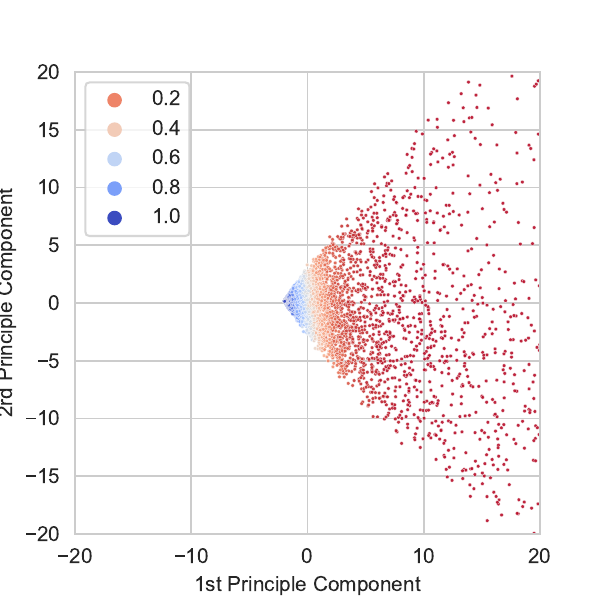}
\caption{Weight Visualization Based on The First and Second Principal Component. Observe that the points are rendered blue to red from the left to right in the x-axis. The points near the origin on the left are unpopular items, and the points are the far right are popular items, due to sparsity of the interaction matrix. Observe the x-axis (the first component) dominates the popularity, as items become more popular when it is closer to the right.}
\label{fig:visualization_first_second}
\end{figure}

This section explains the visualization in detail. We visualize each movie in MovieLens-20M dataset. We first conduct the Principal Component Analysis on the interaction matrix $D$. After PCA decomposition, we plot each movie, with the x-axis and y-axis being the second and third principal components. It is worth noting that the first principal component is omitted as it is proportional to the popularity. The popularity is less related to the semantic similarity; therefore, we instead use the second and third components. The reader interested in visualization using the first component can find such an example in Figure~\ref{fig:visualization_first_second}. 

In Figure~\ref{fig:visualize_weight_comparison}, we visualize the weights proposed from POSIT $\bar{\alpha_i}$, Adversary Score $S_i$ and weights from the best IPW model. The x-axis and the y-axis are their second and third principal components.The close distance between the points indicates that the represented movies are similar. We color each point according to their assigned weight. A cool color represents a larger weight, and a warm color indicates a smaller weight. Overall, the points near the center tend to be unpopular items, as items with sparse interaction are rendered close to zero. Similarly, items away from the center indicate popular ones. In Figure~(a), we show the weights from our method. Notice that our weights focus on semantically close points, but much less on the popular ones. Figure(b) shows the case where we directly use the item advantage score to define the weight without an adversary. We can observe that different colors are mixed everywhere. Observe that some small weights (red points) are given to unpopular items in the center of semantic tails. Compared with Figure~(b), our adversary in Figure~(a) effectively approximate a semantic tail from the advantage score and displays contrasting but progressive colors. In Figure~(c), we demonstrate the weight obtained from the best IPW model. Although it exhibits progressive colors, it does not effectively capture the semantic tails as ours.

\section{Table Including NDCG@20 NDCG@50}
\label{sec:ndcg2050}

% \usepackage{booktabs}
% \usepackage{multirow}
\begin{table}[h]
\resizebox{\columnwidth}{!}{%
\centering
\begin{tabular}{@{}ccccccccc@{}}
\toprule
\multirow{3}{*}{Dataset}       & \multirow{3}{*}{Method} & \multicolumn{4}{c}{Metric}                                                                                      \\ \cmidrule{3-6}
                               &                         & \multicolumn{3}{c}{NDCG@K}   & \multicolumn{3}{c}{Recall@K}                                                         \\ \cmidrule(lr){3-5} \cmidrule(lr){6-8}
                               &                         & \multicolumn{1}{c}{Top100} & \multicolumn{1}{c}{Top50} & \multicolumn{1}{c}{Top20} & \multicolumn{1}{c}{Top100} & \multicolumn{1}{c}{Top50} & \multicolumn{1}{c}{Top20} \\
\midrule
\multirow{7}{*}{Movie Lens}    & \system                    & \textbf{0.4214}  & \textbf{0.3823} & \textbf{0.3402}            & \textbf{0.6369}                & \textbf{0.5244}                & \textbf{0.3928}               \\
                               & IPW                     & \textbf{0.4209}    & \textbf{0.3821}  & \textbf{0.3404}                 & \textbf{0.6364}                         & 0.5226                         & \textbf{0.3926}                        \\
                               & CVar                    & 0.4200    & 0.3810 & \textbf{0.3393}                   & 0.6353                         & 0.5210                          & 0.3911                        \\
                               & Rerank                  & 0.4190    & 0.3806 & 0.3390                   & 0.6300                         & 0.5199                         & 0.3906                        \\
                               & EASE                    & 0.4199    &0.3808 & 0.3390                   & 0.6356                         & 0.5209                         & 0.3906                        \\ 
                               & Causal                & 0.3695    & 0.3340 &0.2966              & 0.5599     &      0.4560   &   0.3411 \\
                               & Post & 0.3927 & 0.3808 & 0.3390 & 0.6339 & 0.5210 & 0.3906\\
                               & MP                    &          0.1905 &0.1598 & 0.1354             &        0.3300                &    0.2351                      &           0.1617              \\                               \cmidrule(lr){1-8}
\multicolumn{2}{c}{Standard Deviation}                 & \multicolumn{6}{c}{$\pm 0.0010$}\\
 \bottomrule
\end{tabular}
}
\caption{An example with NDCG@20 and NDCG@50. We can find they share similar trends.}

\label{tab:tot_performance}

\end{table}

Due to the space limit, we did not include NDCG@ 20,50 in the main text. We empirically found that they convey similar information. Therefore, here we show the result on Movie Lens dataset. We include the results in \autoref{tab:tot_performance}.

\end{document}